\documentclass[fleqn,usenatbib]{mnras}

\usepackage{newtxtext,newtxmath}

\usepackage[T1]{fontenc}

\DeclareRobustCommand{\VAN}[3]{#2}
\let\VANthebibliography\thebibliography
\def\thebibliography{\DeclareRobustCommand{\VAN}[3]{##3}\VANthebibliography}

\usepackage{graphicx}	
\usepackage{amsmath}	

\newcommand{\Velajr}{Vela Junior}
\newcommand{\Msun}{M_\odot}
\newcommand{\XMM}{XMM-Newton}
\newcommand{\NH}{N_{\rm H}}
\newcommand{\Tinfty}{T^\infty}
\newcommand{\Rinfty}{R^\infty}
\newcommand{\Tinftyb}{T_2^\infty}
\newcommand{\Rinftyb}{R_2^\infty}
\newcommand{\Linfty}{L^\infty}
\newcommand{\Teff}{T_{\rm eff}}
\newcommand{\Rem}{R_{\rm em}}
\newcommand{\zg}{z_{\rm g}}

\newcommand{\Tb}{T_2}
\newcommand{\Rb}{R_2}
\newcommand{\Teffb}{T_{\rm eff,2}}
\newcommand{\Remb}{R_{\rm em,2}}
\newcommand{\Eb}{E_B}
\newcommand{\Ebinfty}{\Eb^\infty}

\title[X-ray spectra of the \Velajr\ CCO]{Neutron star cooling implications and magnetic field of the \Velajr\ central compact object from all \XMM\ and Chandra spectra}

\author[Ho, Simkhayeva, and Potekhin]{
Wynn C. G. Ho,$^{1}$\thanks{E-mail: who@haverford.edu}
Esther Simkhayeva,$^{1}$
and
Alexander Y. Potekhin$^{2}$
\\
$^{1}$Department of Physics and Astronomy, Haverford College, 370 Lancaster Avenue, Haverford, PA 19041, USA\\
$^{2}$Ioffe Institute, Politekhnicheskaya 26, Saint Petersburg, 194021, Russia\\
}

\date{Accepted 2026 January 16. Received 2026 January 13; in original form 2025 December 14}

\pubyear{2026}

\begin{document}
\label{firstpage}
\pagerange{\pageref{firstpage}--\pageref{lastpage}}
\maketitle

\begin{abstract}
The central compact object (CCO) in the \Velajr\ supernova remnant is a young
neutron star whose relatively low X-ray flux and small distance suggest it
has a mass high enough to activate fast neutrino cooling processes.
Here we analyse all \XMM\ MOS and pn and Chandra ACIS-S spectra of the
\Velajr\ CCO, with observations taking place over the 9 years from 2001 to 2010.
We find that the best-fit flux and spectral model parameters do not vary
significantly when treating each observation independently,
and therefore we fit all the spectra simultaneously using various spectral
models to characterize the predominantly thermal emission from the neutron
star surface.
Our results indicate the \Velajr\ CCO has an atmosphere composed of hydrogen,
a hot spot temperature (unredshifted) of $3.5\times10^{6}\mbox{ K}$,
and a colder surface temperature of $(6.6-8.8)\times10^{5}\mbox{ K}$.
Possible absorption lines at $\approx$0.6~keV and 0.9~keV provide evidence
for the first-time of an average surface magnetic field
$B\approx3\times10^{10}\mbox{ G}$
for this CCO, which is similar to the magnetic field of other CCOs.
At the accurate new \Velajr\ distance of $1.4$~kpc,
the observed luminosity that is dominated by the hot spot is
$\sim5\times10^{32}\mbox{ erg s$^{-1}$}$.  The luminosity from the rest
of the colder surface is $(1.3-4.0)\times10^{32}\mbox{ erg s$^{-1}$}$.
The cool luminosity and temperature imply the \Velajr\ CCO is indeed
colder than many other young neutron stars and probably has a high mass
that triggered fast neutrino cooling.
\end{abstract}




\section{Introduction} \label{sec:intro}

The cooling behaviour of neutron stars (NSs) is a crucial probe for
uncovering properties of nuclear matter at high densities
(see, e.g., \citealt{potekhinetal15,burgioetal21}, for review).
For example, the rapid real-time cooling measured over more than two decades
of the NS in the Cassiopeia~A supernova remnant (SNR)
can be explained by neutrino emission due to Cooper pair breaking and
formation processes and thus
provides constraints on the critical energy gaps/temperatures of nucleon
superconductivity and superfluidity
\citep{pageetal11,shterninetal11,shterninetal23,zhaoetal25}.
Alternatively, this rapid cooling can be explained by direct Urca
processes, which are more efficient than pair breaking and formation
but only predicted to occur for some nuclear equations of state
\citep{lattimeretal91,pageapplegate92},
and thus can constrain characteristics of the NS core
(see, e.g., \citealt{potekhinyakovlev26}).
Another example is the low observed temperature of a small number of young NSs
(see, e.g., \citealt{pageetal04}),
including the NS in the \Velajr\ SNR \citep{potekhinetal20,marinoetal24},
which also point to fast direct Urca cooling.
Here we re-evaluate the \Velajr\ NS in this context and in light of its
recent distance determination \citep{suherlietal26} and much
more data simultaneously analysed together than done previously.

The \Velajr\ SNR (also known as RX~J0852.0$-$4622 and G266.2$-$1.2)
is young, with an age of 2.4--5.1~kyr \citep{allenetal15}, and
hosts CXOU~J085201.4$-$461753, a NS that was confirmed by Chandra
\citep{pavlovetal01} soon after discovery of the SNR \citep{aschenbach98}.
This NS is one member of a class of NSs known as central compact objects (CCOs).
CCOs are made up of about a dozen young NSs that are located near the centre
of their associated SNR and are nearly only seen to radiate constant thermal
X-ray emission from their surface \citep{deluca08,deluca17,gotthelfetal13}.

CCOs also appear to possess low magnetic fields ($\lesssim10^{11}\mbox{ G}$)
compared to most young pulsars and NSs,
with the main evidence coming from the only three CCOs whose spin period $P$
and time derivative of period $\dot{P}$ have been measured.
Attributing each CCO's spin-down rate $\dot{P}$ to energy loss by
magnetic dipole radiation,
PSR~J0821$-$4300 with $P=112\mbox{ ms}$ in the Puppis~A SNR has a magnetic field
$B=2.9\times10^{10}\mbox{ G}$ \citep{gotthelfhalpern09,gotthelfetal13},
PSR~J1210$-$5226 (also known as 1E~1207.4$-$5209) with $P=424\mbox{ ms}$ has
$B=9.8\times10^{10}\mbox{ G}$ \citep{zavlinetal00,gotthelfetal13},
and PSR~J1852$+$0040 with $P=105\mbox{ ms}$ in the Kesteven~79 SNR has
$B=3.1\times10^{10}\mbox{ G}$ \citep{gotthelfetal05,halperngotthelf10};
note these values of $B$ are the fields at the NS equator.
For comparison, the spectrum of 1E~1207.4$-$5209 shows clear absorption
lines at 0.7 and 1.4~keV that can be attributed to electron cyclotron
resonance and its first harmonic at $B=(7-8)\times10^{10}\mbox{ G}$
\citep{sanwaletal02,mereghettietal02,bignamietal03,suleimanovetal12},
and the spectrum of PSR~J0821$-$4300 appears to have a rotation
phase-dependent feature(s) at 0.46 or 0.75~keV that implies
$B\sim5\times10^{10}\mbox{ G}$ if also interpreted as being due to
electron cyclotron resonance \citep{gotthelfhalpern09,gotthelfetal13}.

\begin{table}
\centering
\caption{Log of observations used in this work.}
\label{tab:data}
\begin{tabular}{lccc}
  \hline
Instrument & ObsID & Observation & Exposure \\
 & & date & (ks) \\
  \hline
Chandra ACIS-S & 1034 & 2001 Sep 17 & 31 \\
\XMM\ MOS & 0112870601 & 2001 Apr 27 & 25 \\
\XMM\ MOS+pn & 0147750101 & 2003 May 21 & 56 \\
\XMM\ MOS+pn & 0147750201 & 2003 Jun 25 & 17 \\
\XMM\ MOS+pn & 0207300101 & 2005 Jun 2 & 54 \\
\XMM\ MOS+pn & 0652510101 & 2010 Nov 13 & 85 \\
\hline
\end{tabular}
\end{table}

\begin{table*}
\renewcommand{\arraystretch}{1.3}
\centering
\caption{Spectral modelling results from previous works using blackbody (BB),
neutron star atmosphere
(\texttt{carbatm}, \texttt{hatm}, \texttt{nsa}, \texttt{nsmaxg}, \texttt{nsx}),
and/or power law (PL) models.
Here, $kT$ (and $k\Tb$) refers to $k\Tinfty$ for a blackbody model or
$k\Teff$ for an atmosphere model and similarly for the radius $R$,
relative to distance $d$.
Errors are 1$\sigma$, except for those from \citet{potekhinetal20,marinoetal24}
which are at 90~percent confidence and those from \citet{alfordhalpern23}
which are at unknown confidence.
See text for more details.}
\label{tab:previous}
\begin{tabular}{lcccccccc}
  \hline
Reference work & Dataset & Spectral & $\NH$ & $kT$ & $R/d$ & $k\Tb$ or $\Gamma$ & $\Rb/d$ & $\chi^2$/dof \\
& & model & ($10^{21}\mbox{ cm$^{-2}$}$) & (eV) & (km/kpc) & (eV) & (km/kpc) & \\
  \hline
\citealt{kargaltsevetal02} & ACIS-S & BB & 3.45$\pm$0.15 & 404$\pm$5 &  0.28$\pm$0.01 & $<89$ & 10 & 84/74 \\
\citealt{kargaltsevetal02} & ACIS-S & \texttt{nsa} & & 270 &  1.2 & & & \\
\citealt{beckeretal06} & MOS+pn 2001+2003 & BB & 3.22$^{+0.14}_{-0.13}$ & 391$\pm$4 & 0.285$^{+0.009}_{-0.008}$ & & & 521/431 \\
\citealt{beckeretal06} & MOS+pn 2001+2003 & BB+BB & 3.82$^{+0.36}_{-0.30}$ & 343$^{+24}_{-40}$ & 0.36$^{+0.05}_{-0.03}$ & 569$^{+261}_{-101}$ & 0.059$^{+0.064}_{-0.043}$ & 465/429 \\
\citealt{beckeretal06} & MOS+pn 2001+2003 & BB+PL & 7.63$^{+0.72}_{-1.59}$ & 366$^{+10}_{-13}$ & 0.30$^{+0.03}_{-0.02}$ & 4.21$^{+0.30}_{-0.59}$ & & 469/429 \\
\citealt{potekhinetal20}$^a$ & MOS+pn 2003--2010 & \texttt{nsx} [carbon] & 6.8$\pm$0.1 & 133$\pm$2 & 5.0$\pm$0.4 & & & 920/913 \\
\citealt{wuetal21} & pn 2010 & BB+BB & 5.5$^{+0.6}_{-0.5}$ & 317$^{+33}_{-44}$ & 0.39$^{+0.08}_{-0.04}$ & 482$^{+89}_{-43}$ & 0.12$^{+0.07}_{-0.06}$ & 59/68 \\
\citealt{wuetal21} & pn 2010 & \texttt{hatm}+\texttt{hatm} & 7.96$^{+0.67}_{-1.10}$ & 256$^{+6}_{-5}$ & 0.95 & 71$^{+5}_{-9}$ & 12 & 61/69 \\
\citealt{wuetal21} & pn 2010 & \texttt{carbatm} & 7.0$\pm$0.2 & 145$\pm$3 & 4.3 & & & 60/70 \\
\citealt{alfordhalpern23} & pn 2005 & \texttt{carbatm} & 4.70$\pm$0.12 & 132$\pm$0.4 & 5.2 & & & 82/73 \\
\citealt{alfordhalpern23} & pn 2010 & \texttt{carbatm} & 4.91$\pm$0.11 & 132$\pm$0.3 & 5.2 & & & 87/80 \\
\citealt{marinoetal24} & ACIS-S & BB & 4.7$\pm$0.4 & 397$\pm$9 &  0.24$\pm$0.08 & $<50$ & 11 & 162/143 \\
\citealt{marinoetal24} & ACIS-S & \texttt{nsmaxg} [$10^{12}\mbox{ G}$] & & 310$\pm$10 & 0.50$^{+0.07}_{-0.05}$ & & & \\
\hline
\multicolumn{9}{l}{$^a$ P.~Shternin, private comm.}
\end{tabular}
\end{table*}

After the \Velajr\ CCO
(hereafter, we refer for simplicity to the NS/CCO as \Velajr)
was confirmed and its position of
R.A.~$=08^{\rm h}52^{\rm m}01^{\rm s}\!\!.38$,
decl.~$=-46^\circ17\arcmin53\farcs34$ (J2000) was derived
using a short 3~ks Chandra ACIS-I observation in 2000
(ObsID 1032; \citealt{pavlovetal01}),
 \Velajr\ was observed twice more with Chandra.
These two are a 31~ks observation in 2001 using ACIS-S in continuous
clocking mode (ObsID 1034) and a 28~ks observation in 2009 using
HRC-I (ObsID 10702).
While the ACIS-S observation was used to study the spectrum of \Velajr\
\citep{kargaltsevetal02,marinoetal24},
the HRC-I observation was used along with the ACIS-I observation to
derive a 3$\sigma$ upper limit of 300~mas~yr$^{-1}$ on the proper motion
of \Velajr\ (\citealt{mignanietal19}; see also \citealt{camillonietal23}).
Meanwhile, \XMM\ observed \Velajr\ five times from 2001 to 2010,
as listed in Table~\ref{tab:data}.
Various works used different subsets of the \XMM\ observations to study the
spectrum of \Velajr, fitting it with various single or multi-component models
\citep{beckeretal06,potekhinetal20,wuetal21,alfordhalpern23}.
A summary of the main spectral results from these previous analyses is
given in Table~\ref{tab:previous}.

In this work, we analyse all the \XMM\ data and the Chandra ACIS-S
data together for the first time, and we conduct fits to the spectra
systematically using different models.
An outline of the paper is as follows.
Section~\ref{sec:data} describes the Chandra and \XMM\ data
analysed in this work and their processing.
Section~\ref{sec:results} presents our results.
Section~\ref{sec:discuss} summarizes our findings and discusses their
implications for NS cooling and dense nuclear matter.

\section{Data analysis} \label{sec:data}

\subsection{Chandra data} \label{sec:chandra}

Chandra observed \Velajr\ using the ACIS-S detector in continuous clocking
(CC) mode on 2001 September 17 (ObsID 1034) for 31~ks.
Meanwhile, we ignore the short 3~ks ACIS-I observation from 2000 October 26
(ObsID 1032) and the poorer spectral resolution 28~ks HRC-I observation from
2009 November 13 (ObsID 10702).
We reprocess the ACIS-S data following the standard procedure using
\texttt{chandra\_repro} of the Chandra Interactive Analysis of
Observations (CIAO) package version 4.17 and Calibration Database
(CALDB) 4.12.0 \citep{fruscioneetal06}.
We follow the recommended procedure for extracting source events from CC mode
data\footnote{\url{https://cxc.cfa.harvard.edu/ciao/caveats/acis_cc_mode.html}}.
Using \texttt{specextract}, source events are extracted from
a rotated 6$\times$4~pixel box centred on the CCO position,
while two rotated 14$\times$4~pixel boxes on either side are used for
background.
Spectra are binned using \texttt{dmgroup} with a minimum of 100 counts per bin.

\subsection{\XMM\ data} \label{sec:xmm}

\XMM\ EPIC observed \Velajr\ on five occasions, as shown in
Table~\ref{tab:data}.
The first observation in 2001 (ObsID 0112870601) was pointed
more than 2~arcmin off-axis from \Velajr;
furthermore, the source fell on a gap on the pn detector,
and therefore we ignore this pn data.
All MOS observations are in full frame mode, while pn observations are
in small window mode.
We process MOS and pn observation data files (ODFs) using the
Science Analysis Software (SAS) 22.0.0 tasks \texttt{emproc} and
\texttt{epproc}, respectively.
To remove periods of background flaring, we extract single event
($\mbox{PATTERN}=0$), high energy ($>10$~keV for MOS and $10-12$~keV for pn)
light curves from which we determine low and steady count rate thresholds
of $0.8-2.5\mbox{ counts s$^{-1}$}$ for MOS and
$0.1-0.2\mbox{ counts s$^{-1}$}$ for pn, depending on the observation.
These thresholds are used to generate good time intervals (GTIs) with
\texttt{tabgtigen}, and the GTIs are then used to produce flare-cleaned
events with \texttt{evselect}.
The resulting data have effective exposure times
for ObsID 0112870601 of 8~ks each for MOS1 and MOS2,
for ObsID 0147750101 of 38~ks each for MOS1 and MOS2 and 28~ks for pn,
for ObsID 0147750201 of 15~ks each for MOS1 and MOS2 and 11~ks for pn,
for ObsID 0207300101 of 49~ks each for MOS1 and MOS2 and 36~ks for pn,
for ObsID 0652510101 of 70~ks for MOS1, 69~ks for MOS2, and 51~ks for pn.

To extract MOS and pn source counts, we use circular regions with radii of
24~arcsec and 20~arcsec, respectively.
To determine the background, we use annuli with inner and outer radii of
25~arcsec and 40~arcsec for MOS and 22~arcsec and 40~arcsec for pn.
Using \texttt{epatplot}, we do not find the source counts to be significantly
affected by pile-up.
Source and background counts for spectral analysis are extracted
using $\mbox{PATTERN}\le12$ and $\le4$ for MOS and pn, respectively,
and $\mbox{FLAG}=0$ for spectra.
We calculate source and background areas and account for bad pixels and
chip gaps using \texttt{backscale}.
We then compute rmf and arf files.
MOS1 and MOS2 spectra are combined using \texttt{epicspeccombine}.
The two 2003 observations, ObsIDs 0147750101 and 0147750101,
are also combined since they took place only about one month apart.
Spectra are binned using \texttt{specgroup} to a minimum of 100 counts
per bin for the combined MOS spectrum and for the pn spectrum.

\subsection{Spectral modelling} \label{sec:models}

We perform spectral fitting using Xspec 12.15.0 \citep{arnaud96}
and consider the energy range 0.5$-$8~keV.
The spectral models used here are composed of several components.
When fitting across instruments, we include the \texttt{constant}
model, in order to account for instrumental differences between ACIS-S,
MOS, and pn spectral normalizations, and we fix its value to 1 for pn
spectra and allow it to vary for the ACIS-S spectrum and the combined
MOS spectra.
Next, we include a component to account for photoelectric absorption
by the interstellar medium, i.e. \texttt{tbabs} with abundances from
\citet{wilmsetal00} and cross-sections from \citet{verneretal96} and
parametrized by the effective hydrogen column density $\NH$.

The final component(s) used, which models the intrinsic spectrum of \Velajr,
is either a blackbody (BB; \texttt{bbodyrad}), a NS atmosphere,
or a power law (PL; \texttt{powerlaw}).
The blackbody is parametrized by the redshifted temperature $k\Tinfty$
and normalization $\Rinfty/d$, where $\Tinfty=T/(1+\zg)$, $\Rinfty=R(1+\zg)$,
distance is $d$, redshift factor is $1+\zg=1/(1-2GM/c^2R)^{1/2}$,
and NS mass is $M$ and radius is $R$.
The NS atmosphere models we use
(\texttt{nsx} and \texttt{nsmaxg}; see below)
have similar parameters, in particular, effective temperature $k\Teff$, $M$,
$R$, $d$, and normalization $\Rem/R$, where $\Rem$ is emission region radius,
as well as atmosphere composition and, in the case of \texttt{nsmaxg},
magnetic field strength $B$.
Previous works showed that a single power law
is a poor fit to \Velajr\ spectra \citep{kargaltsevetal02,beckeretal06},
and we also find this to be the case.
However, we do consider models where a power law is used in combination
with a blackbody or atmosphere.
The power law is parametrized by photon index $\Gamma$ and normalization.
In summary, we model the intrinsic \Velajr\ spectrum by either a single
component made up of a blackbody or atmosphere spectrum that represents
a hot spot on the NS surface
or two components made up of two blackbodies (BB+BB)
or two atmospheres, a blackbody and a power law (BB+PL), or an atmosphere
and a power law.  For the two (temperature) atmospheres case, the second
(cooler) atmosphere has a fixed normalization $\Rem/R=1$ to represent
emission from the rest of the NS surface besides the hot spot.
Once a best-fit is found, we use \texttt{cflux} to calculate flux
and its uncertainties while holding all other model parameters at
their best-fit values.

For NS atmosphere models, we consider two types of models, depending
on the unknown surface magnetic field of \Velajr.
The first are those computed for partially ionized hydrogen or carbon
in the low/non-magnetic regime when $B\lesssim10^9\mbox{ G}$
(\texttt{nsx}; \citealt{hoheinke09}).
A non-magnetic helium atmosphere (via \texttt{nsx}) yields fits that are
comparable to those of hydrogen and carbon and give model parameter values
in between those of hydrogen and carbon.
However, a helium or carbon atmosphere is unlikely to be present for
isolated NSs at the age and temperature of \Velajr\
\citep{changetal10,wijngaardenetal19}, and thus we do not show our
helium results.
On the other hand, we show our carbon results so that they can be
compared to results from previous works (see Table~\ref{tab:previous}).
The second type of models are those for partially ionized hydrogen
at a single $B$ in the range $10^{10}\mbox{ G}\le B\le3\times10^{13}\mbox{ G}$
(\texttt{nsmaxg}\footnote{\url{https://heasarc.gsfc.nasa.gov/docs/software/xspec/manual/node208.html}}; \citealt{hoetal08,ho14,potekhinetal14}).
We note that the four atmosphere models with varying magnetic fields and
surface temperatures in \texttt{nsmaxg} do not fit the data well when used
as the only component for the intrinsic \Velajr\ spectrum, but they can
fit the data when used in combination with a power law.
Meanwhile, partially ionized carbon, oxygen, and neon atmosphere models at
$10^{12}\mbox{ G}$ and $10^{13}\mbox{ G}$ (via \texttt{nsmaxg};
\citealt{moriho07}) also do not fit the data, so will not be discussed further.
For all atmosphere spectral fits, model parameters for NS mass and radius
are fixed to $M=1.4\,\Msun$ and $R=12\mbox{ km}$, respectively;
as a consequence, a factor $1+\zg=1.235$ is used to
convert between redshifted and unredshifted values when needed.
We note that a small change in the fixed mass, e.g.,
to $1.6\,\Msun$, and hence redshift factor, leads to small differences
in best-fit values that are mostly within the uncertainties.

\section{Results} \label{sec:results}

\begin{table}
\renewcommand{\arraystretch}{1.3}
\centering
\caption{Spectral modelling results using \texttt{bbodyrad} (BB),
where all parameters are free to vary between datasets.
Absorbed 0.5--8~keV flux $f_{0.5-8}^{\rm abs}$ is in units of
$10^{-12}$~erg~cm$^{-2}$~s$^{-1}$.
Errors are 1$\sigma$.
For the simultaneous fit of all datasets shown in the last row, two additional
parameters are included in the fit: a constant MOS/pn normalization and a
constant ACIS-S/pn normalization; the best-fit for these two are
$0.969\pm0.006$ and $1.12\pm0.01$, respectively.}
\label{tab:bb}
\begin{tabular}{cccccc}
  \hline
Dataset & $\NH$ & $k\Tinfty$ & $\Rinfty/d$ & $f_{0.5-8}^{\rm abs}$ & $\chi^2$/dof \\
& ($10^{21}\mbox{ cm$^{-2}$}$) & (eV) & (m/kpc) & & \\
  \hline
ACIS-S & 4.13$^{+0.21}_{-0.20}$ & 404$^{+5}_{-5}$ & 278$^{+9}_{-9}$ & 1.42$\pm$0.01 & 86/73 \\
MOS 2001 & 4.08$^{+0.59}_{-0.56}$ & 397$^{+15}_{-14}$ & 269$^{+29}_{-23}$ & 1.24$\pm$0.03 & 22/22 \\
MOS 2003 & 4.31$^{+0.20}_{-0.19}$ & 393$^{+4}_{-4}$ & 277$^{+8}_{-7}$ & 1.22$\pm$0.01 & 84/61 \\
MOS 2005 & 4.30$^{+0.22}_{-0.22}$ & 400$^{+4}_{-4}$ & 264$^{+8}_{-8}$ & 1.21$\pm$0.01 & 60/57 \\
MOS 2010 & 4.30$^{+0.17}_{-0.17}$ & 395$^{+3}_{-3}$ & 273$^{+7}_{-6}$ & 1.217$\pm$0.009 & 87/66 \\
 pn 2003 & 4.29$^{+0.16}_{-0.15}$ & 396$^{+4}_{-4}$ & 271$^{+7}_{-7}$ & 1.23$\pm$0.01 & 71/62 \\
 pn 2005 & 4.07$^{+0.17}_{-0.16}$ & 396$^{+4}_{-4}$ & 275$^{+8}_{-7}$ & 1.27$\pm$0.01 & 74/54 \\
 pn 2010 & 4.40$^{+0.14}_{-0.14}$ & 395$^{+3}_{-3}$ & 278$^{+6}_{-6}$ & 1.259$\pm$0.009 & 62/59 \\
All & 4.27$^{+0.07}_{-0.06}$ & 396$^{+1}_{-1}$ & 275$^{+3}_{-3}$ & 1.246$\pm$0.004 & 573/472 \\
\hline
\end{tabular}
\end{table}

\begin{figure}
\includegraphics[width=\columnwidth]{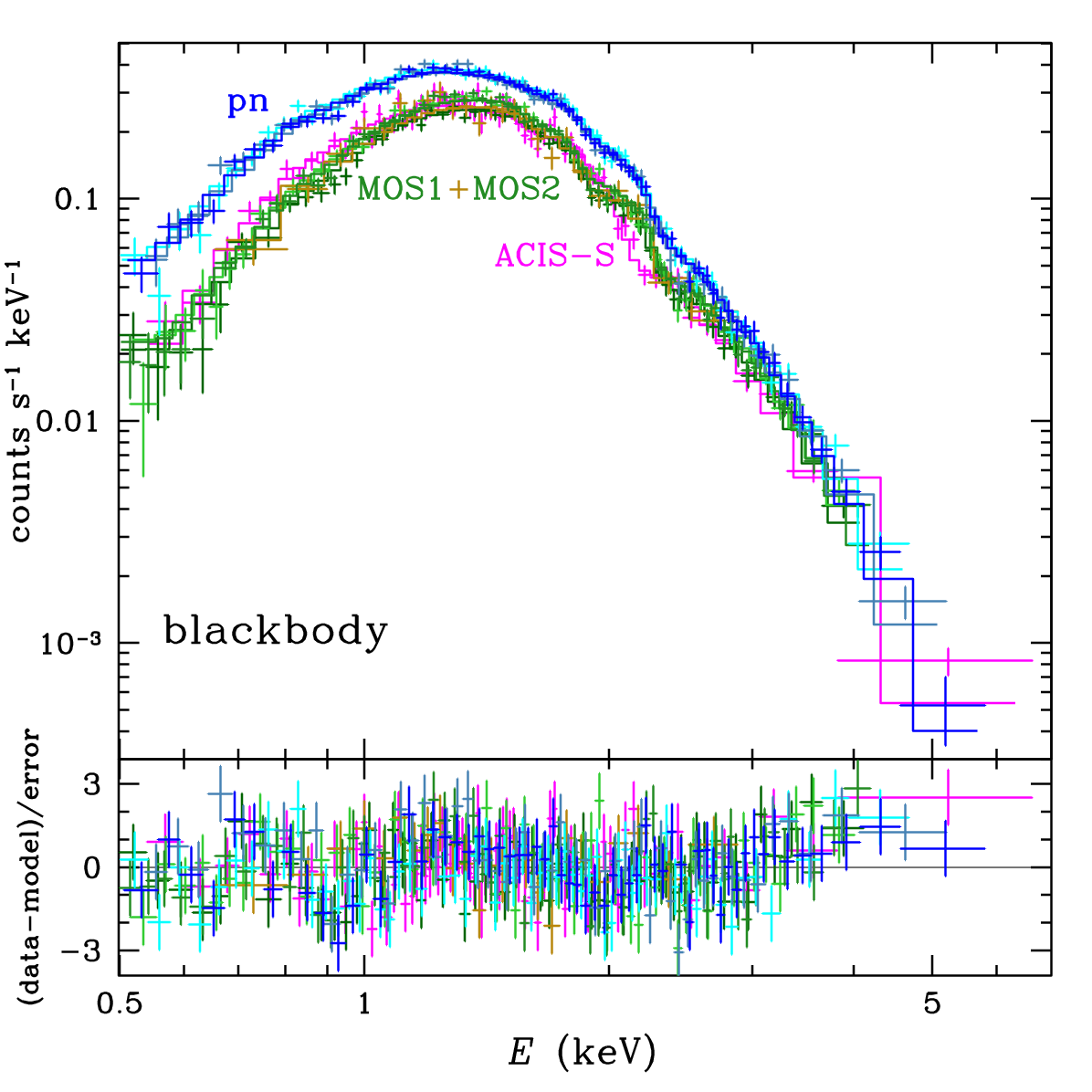}
\caption{
Spectra of \Velajr\ from Chandra ACIS-S and \XMM\ MOS1+MOS2 and pn data.
Top panel shows data with 1$\sigma$ errors (crosses) and
spectral model made up of a single blackbody.
Bottom panel shows $\chi^2=\mbox{(data-model)/error}$.
}
\label{fig:bb}
\end{figure}

We start by fitting each dataset with a single component absorbed blackbody.
Our best-fit results are given in Table~\ref{tab:bb}.
Individually, some of the datasets can be fit well with a blackbody,
while others are not well-fit, and they all indicate emission primarily
from a small ($\Rinfty\sim0.3\mbox{ km}$ at 1~kpc) hot spot on the NS surface.
The quality of the fit depends in large part on whether the spectra at
$\gtrsim 3\mbox{ keV}$ are taken into account.
For example, the MOS spectrum from 2001
(from a 8~ks effective exposure taken off-axis; see Section~\ref{sec:xmm})
spans only 0.7--3~keV with our binning.
Even though the flux at $>3\mbox{ keV}$ is low, a single blackbody
at $k\Tinfty\approx400\mbox{ eV}$ fails to characterize this high-energy
emission, as noted by \citet{beckeretal06}.
We find a very similar temperature and radius when comparing our
results to previous works that fit single blackbodies to specific datasets,
while our best-fit $\NH\sim4.3\times10^{21}\mbox{ cm$^{-2}$}$
is somewhat different from the results of \citet{kargaltsevetal02,beckeretal06}
but is consistent with that of \citet{marinoetal24}
(see Table~\ref{tab:previous}).
Finally, from Table~\ref{tab:bb} and as we shall show below, the spectra
do not appear to vary significantly from 2001 to 2010, and therefore we
fit all the datasets together with one absorbed blackbody.
The results are given in Table~\ref{tab:bb} and shown in Figure~\ref{fig:bb}.
We see the fit quality is not good ($\chi^2/\mbox{ dof}=573/472=1.21$),
and the model fit residual shows wave-like systematics and underprediction
of the data at $>3\mbox{ keV}$.

\begin{table*}
\renewcommand{\arraystretch}{1.3}
\centering
\caption{Spectral modelling results using \texttt{bbodyrad}+\texttt{bbodyrad}
(BB+BB), where all parameters are free to vary between datasets.
Absorbed 0.5--8~keV flux $f_{0.5-8}^{\rm abs}$ is in units of
$10^{-12}$~erg~cm$^{-2}$~s$^{-1}$.
Errors are 1$\sigma$.
For the simultaneous fit of all datasets shown in the last row, two additional
parameters are included in the fit: a constant MOS/pn normalization and a
constant ACIS-S/pn normalization; the best-fit for these two are
$0.968\pm0.006$ and $1.12\pm0.01$, respectively.}
\label{tab:2bb}
\begin{tabular}{cccccccc}
  \hline
Dataset & $\NH$ & $k\Tinfty$ & $\Rinfty/d$ & $k\Tinftyb$ & $\Rinftyb/d$ & $f_{0.5-8}^{\rm abs}$ & $\chi^2$/dof \\
& ($10^{21}\mbox{ cm$^{-2}$}$) & (eV) & (m/kpc) & (eV) & (m/kpc) & & \\
  \hline
ACIS-S & 4.35$^{+0.29}_{-0.22}$ & 395$^{+7}_{-22}$ & 292$^{+21}_{-12}$ & 1700$^{+0}_{-1200}$ & 3$^{+60}_{-3}$ & 1.44$\pm$0.01 & 82/72 \\
MOS 2001 & 13.9$^{+3.5}_{-4.2}$ & 381$^{+46}_{-29}$ & 342$^{+118}_{-83}$ & 127$^{+41}_{-19}$ & 8600$^{+13000}_{-3000}$ & 1.24$\pm$0.03 & 15/20 \\
MOS 2003 & 5.14$^{+0.42}_{-0.37}$ & 347$^{+23}_{-38}$ & 342$^{+35}_{-28}$ & 577$^{+363}_{-111}$ & 56$^{+89}_{-26}$ & 1.23$\pm$0.01 & 67/59 \\
MOS 2005 & 5.11$^{+0.58}_{-0.41}$ & 345$^{+35}_{-42}$ & 319$^{+34}_{-27}$ & 514$^{+418}_{-82}$ & 86$^{+180}_{-42}$ & 1.21$\pm$0.01 & 50/55 \\
MOS 2010 & 5.06$^{+0.74}_{-0.43}$ & 348$^{+33}_{-88}$ & 330$^{+55}_{-27}$ & 535$^{+1600}_{-274}$ & 71$^{+270}_{-35}$ & 1.225$\pm$0.009 & 70/64 \\
 pn 2003 & 4.60$^{+0.22}_{-0.20}$ & 379$^{+8}_{-10}$  &  296$^{+15}_{-12}$ & 783$^{+246}_{-156}$ & 15$^{+28}_{-6}$ & 1.24$\pm$0.01 & 61/60 \\
 pn 2005 & 4.47$^{+0.30}_{-0.24}$ & 370$^{+16}_{-36}$ &  305$^{+20}_{-15}$ & 571$^{+282}_{-126}$ & 46$^{+156}_{-22}$ & 1.28$\pm$0.01 & 66/52 \\
 pn 2010 & 4.79$^{+0.28}_{-0.21}$ & 363$^{+22}_{-52}$ &  302$^{+16}_{-40}$ & 497$^{+192}_{-71}$ & 82$^{+200}_{-38}$ & 1.264$\pm$0.009 & 55/57 \\
All & 4.76$^{+0.12}_{-0.11}$ & 363$^{+9}_{-13}$ &  311$^{+8}_{-7}$ & 548$^{+71}_{-50}$ & 58$^{+44}_{-19}$ & 1.254$\pm$0.004 & 508/470 \\
\hline
\end{tabular}
\end{table*}

\begin{table*}
\renewcommand{\arraystretch}{1.3}
\centering
\caption{Spectral modelling results using \texttt{bbodyrad}+\texttt{powerlaw}
(BB+PL), where all parameters are free to vary between datasets.
Absorbed 0.5--8~keV flux $f_{0.5-8}^{\rm abs}$ is in units of
$10^{-12}$~erg~cm$^{-2}$~s$^{-1}$.
Errors are 1$\sigma$.
For the simultaneous fit of all datasets shown in the last row, two additional
parameters are included in the fit: a constant MOS/pn normalization and a
constant ACIS-S/pn normalization; the best-fit for these two are
$0.969\pm0.006$ and $1.12\pm0.01$, respectively.}
\label{tab:bbpl}
\begin{tabular}{cccccccc}
  \hline
Dataset & $\NH$ & $k\Tinfty$ & $\Rinfty/d$ & $\Gamma$ & PL norm & $f_{0.5-8}^{\rm abs}$ & $\chi^2$/dof \\
& ($10^{21}\mbox{ cm$^{-2}$}$) & (eV) & (m/kpc) & & ($10^{-4}$~cm$^{-2}$~s$^{-1}$~keV$^{-1}$) & & \\
  \hline
ACIS-S & 4.4$^{+4.6}_{-0.3}$ & 394$^{+7}_{-19}$ & 292$^{+23}_{-7}$ & 1.4$^{+3.7}_{-1.4}$ & 0.1$^{+9.3}_{-0.1}$ & 1.44$\pm$0.02 & 82/72 \\
MOS 2001 & 18.0$^{+4.4}_{-4.7}$ &  393$^{+25}_{-24}$ & 305$^{+120}_{-74}$ & 7.1$^{+1.5}_{-1.7}$ & 70$^{+97}_{-49}$ & 1.24$\pm$0.03 & 16/22 \\
MOS 2003 &  8.0$^{+1.7}_{-3.3}$ &  362$^{+9}_{-8}$ & 306$^{+21}_{-22}$ & 3.5$^{+0.5}_{-2.6}$ & 6.3$^{+5.0}_{-6.1}$ & 1.26$\pm$0.01 & 68/59 \\
MOS 2005 &  9.3$^{+1.2}_{-2.2}$ & 377$^{+10}_{-9}$ & 277$^{+20}_{-18}$ & 3.94$^{+0.52}_{-0.65}$ & 9.2$^{+5.5}_{-5.2}$ & 1.23$\pm$0.01 & 48/55 \\
MOS 2010 & 10.0$^{+1.2}_{-1.3}$ &  375$^{+9}_{-8}$ & 280$^{+17}_{-15}$ & 4.15$^{+0.38}_{-0.39}$ & 12$^{+4}_{-4}$ & 1.239$\pm$0.009 & 66/64 \\
 pn 2003 &  9.0$^{+0.9}_{-1.0}$ &  378$^{+5}_{-5}$ & 291$^{+10}_{-10}$ & 4.19$^{+0.26}_{-0.27}$ & 8.3$^{+2.6}_{-2.5}$ & 1.24$\pm$0.01 & 56/60 \\
 pn 2005 &  7.2$^{+1.3}_{-1.5}$ &  377$^{+5}_{-5}$ & 301$^{+9}_{-8}$ & 3.80$^{+0.38}_{-0.48}$ & 4.5$^{+2.9}_{-2.5}$ & 1.29$\pm$0.01 & 63/52 \\
 pn 2010 &  8.0$^{+1.3}_{-2.2}$ &  377$^{+4}_{-4}$ & 305$^{+8}_{-7}$ & 4.05$^{+0.37}_{-0.67}$ & 5.4$^{+3.3}_{-3.8}$ & 1.269$\pm$0.009 & 54/57 \\
All &  8.6$^{+0.5}_{-0.6}$ &  377$^{+2}_{-2}$ & 295$^{+4}_{-4}$ & 4.02$^{+0.15}_{-0.16}$ & 7.6$^{+1.4}_{-1.4}$ & 1.263$\pm$0.004 & 502/470 \\
\hline
\end{tabular}
\end{table*}

\begin{figure}
\includegraphics[width=\columnwidth]{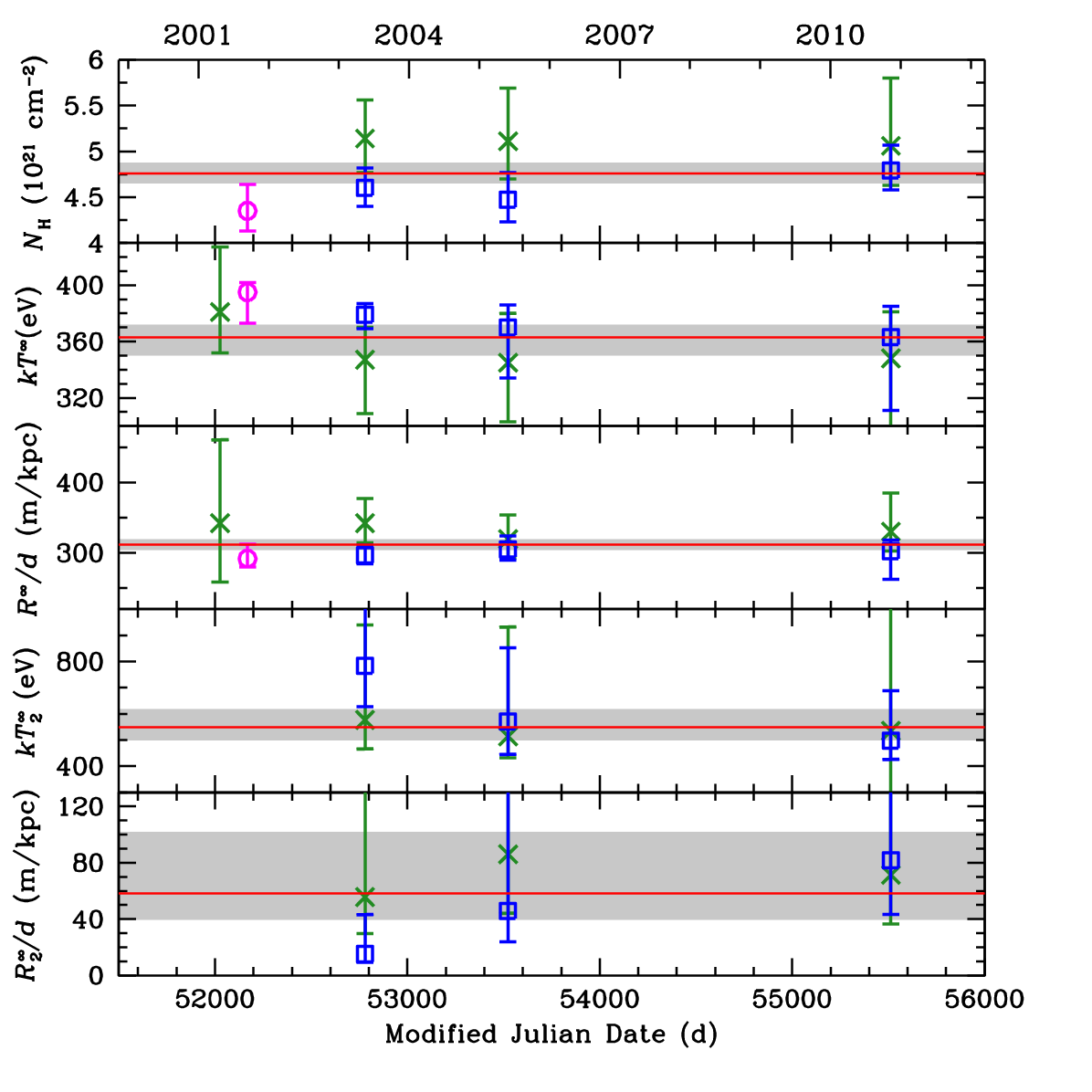}
\caption{
Interstellar absorption $\NH$ and blackbody temperatures $k\Tinfty$ and
$k\Tinftyb$ and emission radii $\Rinfty$ and $\Rinftyb$ (relative to
distance $d$) for a \texttt{bbodyrad}+\texttt{bbodyrad} (BB+BB) model fit
to ACIS-S (circles), MOS (crosses), and pn (squares) spectra.
Errors are 1$\sigma$.
Some values for the ACIS-S and MOS 2001 fits are not shown because they
lie outside displayed ranges due to these data being of lower quality
(see text for details).
Horizontal solid lines show results for a simultaneous fit to all the data,
and the shaded regions encompass the 1$\sigma$ error.
}
\label{fig:2bb}
\end{figure}

\begin{figure}
\includegraphics[width=\columnwidth]{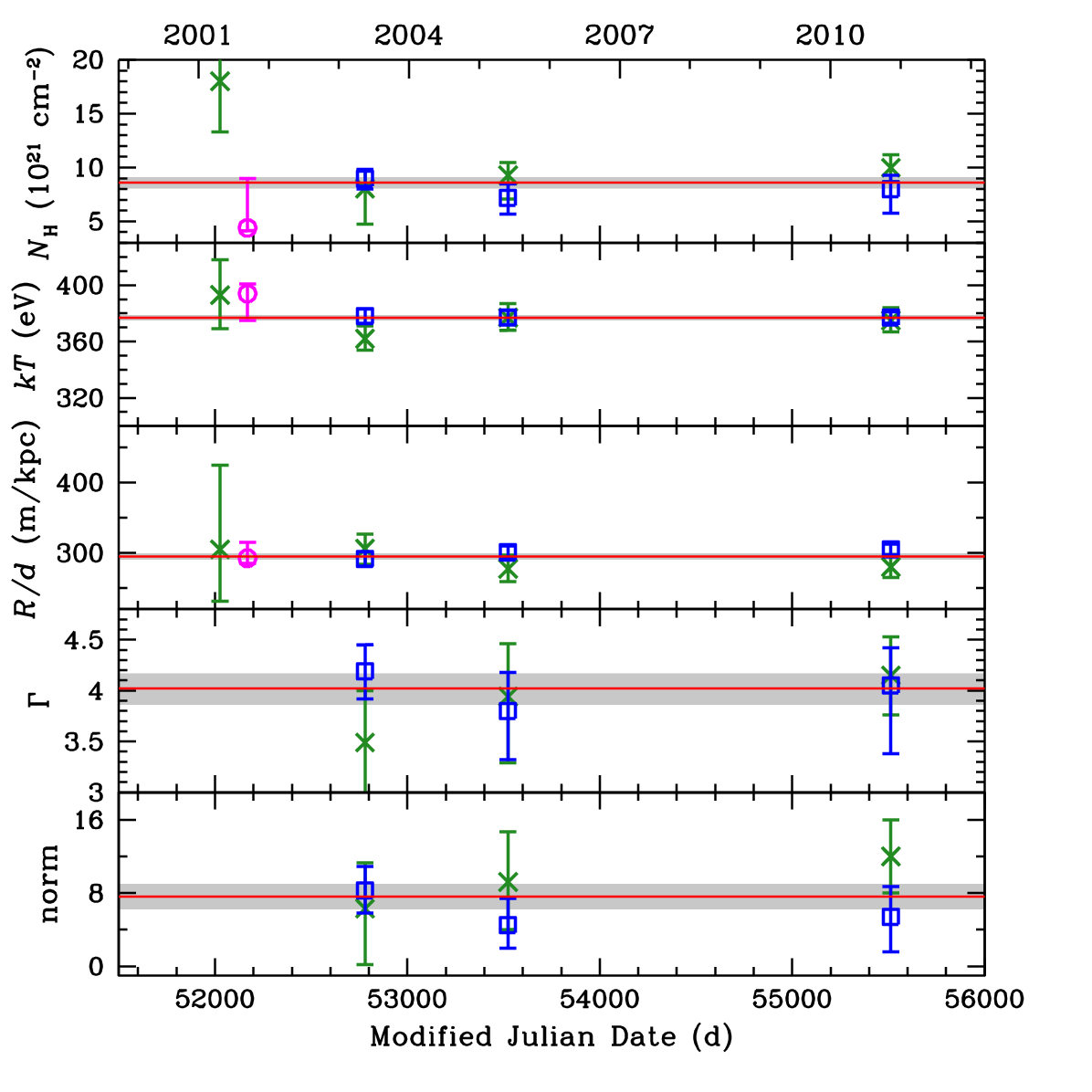}
\caption{
Interstellar absorption $\NH$, blackbody temperature $k\Tinfty$ and
emission radius $\Rinfty$ (relative to distance $d$), and power law index
and normalization for a \texttt{bbodyrad}+\texttt{powerlaw} (BB+PL) model fit
to ACIS-S (circles), MOS (crosses), and pn (squares) spectra.
Errors are 1$\sigma$.
Some values for the ACIS-S and MOS 2001 fits are not shown because they
lie outside displayed ranges due to these data being of lower quality
(see text for details).
Horizontal solid lines show results for a simultaneous fit to all the data,
and the shaded regions encompass the 1$\sigma$ error.
}
\label{fig:bbpl}
\end{figure}

To better characterize the spectra of \Velajr, we fit each dataset
using a multi-component absorbed double blackbody (BB+BB) or blackbody
plus power law (BB+PL).
The best-fit results are shown in
Tables~\ref{tab:2bb} and \ref{tab:bbpl}
and Figures~\ref{fig:2bb} and \ref{fig:bbpl}.
First, note that the addition of a second component, whether a blackbody
or power law, does not significantly improve the goodness-of-fit of the
ACIS-S spectrum,
and thus this extra component is not particularly justified by the data.
In fact, such a conclusion is evident by the best-fit parameters of the
additional component which has a poorly constrained temperature and small
emission radius for the second blackbody and a low normalization for the
power law and hence a very small contribution to the spectrum.
Also of note in the best-fit to the MOS 2001 spectrum is its poorly
constrained and unrealistic parameter values for the extra component and
$\NH$, which is due to the spectrum's limited energy range, as noted above.
Comparing the remaining results to those from the single blackbody fits,
the addition of a second component leads to a decrease of 20---50~eV in
the blackbody temperature and a small increase of the emission radius
(of the cooler blackbody).
In addition, for the double blackbody fits, one can see that the temperature
(emission radius) of the cooler blackbody for the MOS spectra is consistently
lower (higher), but within the uncertainties, than that for the pn spectra;
this is likely due to the lower sensitivity, and hence flux normalization,
of MOS compared to pn.
Our results are also generally consistent with those from previous works
(see Table~\ref{tab:previous}, in particular, \citealt{beckeretal06,wuetal21}).
We note that if one of the two blackbody radii is fixed to $\sim12\mbox{ km}$,
which is much larger than either best-fit emission radii when they are both
allowed to vary (see Table~\ref{tab:2bb}), then the corresponding temperature
is so low that this blackbody component does not contribute significantly
to the best-fit model and the results are essentially the same as those
of a single blackbody model (see Table~\ref{tab:bb}).

It is again evident that the best-fit model parameters do not differ
significantly between the individual observations from 2001 to 2010,
except in some cases for the ACIS-S and MOS data from 2001 due to the
former being taken in CC mode (and thus is partly contaminated by
SNR emission due to data collection being collapsed in one spatial dimension)
and the off-axis short exposure of the latter.
Therefore we fit all the datasets together with either an absorbed double
blackbody or an absorbed blackbody plus power law, with the best-fit results
also shown in Tables~\ref{tab:2bb} and \ref{tab:bbpl}
and Figures~\ref{fig:2bb} and \ref{fig:bbpl}.
We see that both multi-component spectral models provide decent fits to
the data and are much better fits than the single blackbody model.
Both multi-component model fits are comparable, with
$\chi^2/\mbox{dof}=508/470=1.08$ for the BB+BB model being slightly worse
than the $\chi^2/\mbox{dof}=502/470=1.07$ for the BB+PL model.
The absorption $\NH=8.6^{+0.5}_{-0.6}\times10^{21}\mbox{ cm$^{-2}$}$ for BB+PL
[see also the BB+PL model of \citet{beckeretal06} in Table~\ref{tab:previous}]
is higher than for purely thermal models and is on the upper end of the values
$(1-9)\times10^{21}\mbox{ cm$^{-2}$}$ determined for the rims of the SNR
but is still below the $11\times10^{21}\mbox{ cm$^{-2}$}$ determined for
diffuse emission in the SNR central region by
\citet{slaneetal01,aceroetal13,camillonietal23}.
Also while the power law index $\Gamma=4$ is somewhat high,
\citet{beckeretal06} argued that these results by themselves do not mean
the blackbody plus power law can be rejected.

\begin{figure}
\includegraphics[width=\columnwidth]{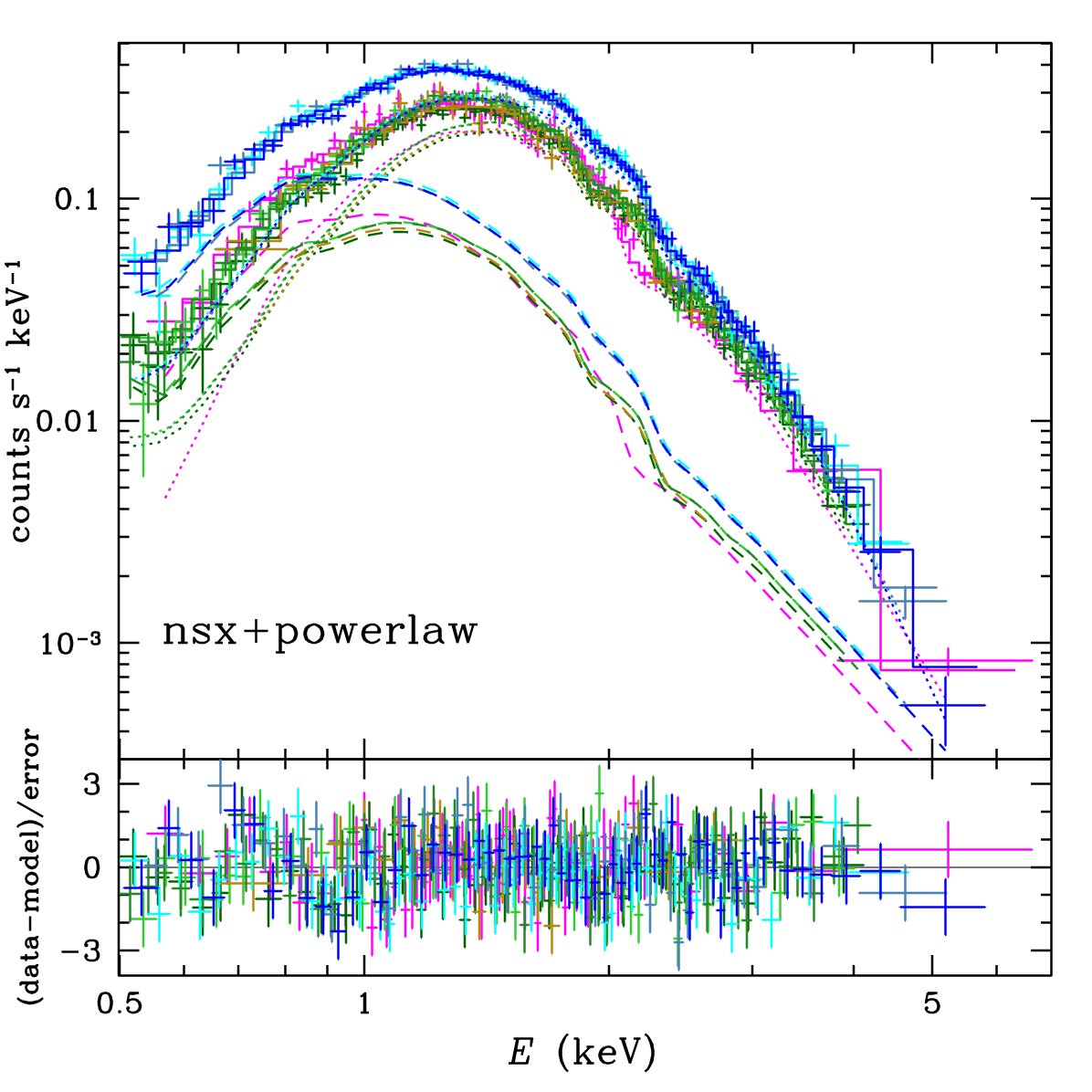}
\caption{
Spectra of \Velajr\ from Chandra ACIS-S and \XMM\ MOS1+MOS2 and pn data.
Top panel shows data with 1$\sigma$ errors (crosses) and
spectral model (\texttt{nsx}+\texttt{PL}) made up of a partially ionized
non-magnetic hydrogen atmosphere component (dotted lines) and
a power law component (dashed lines).
Bottom panel shows $\chi^2=\mbox{(data-model)/error}$.
}
\label{fig:nsxpl}
\end{figure}

We can disfavour the blackbody plus power law model however by examining
more closely the contribution of the power law to modelling the spectra.
A power law model is often used to characterize non-thermal emission at
$>1\mbox{ keV}$ from, e.g., a pulsar wind nebula.
Figure~\ref{fig:nsxpl} shows the simultaneous fit to all the spectra using
a model composed of a NS atmosphere and power law (\texttt{nsx}+\texttt{PL}),
which will be described in more detail below.
In particular, the top panel displays the individual contributions of the
atmosphere component and the power law component to mimicking the actual data.
Our blackbody plus power law (\texttt{BB}+\texttt{PL}) spectral fits show
very similar behaviour.
What is particularly worth highlighting is that the power law component,
while providing some significant flux at $E\gtrsim3\mbox{ keV}$, is the
dominant spectral component at energies below 1~keV.
This also explains the high inferred absorption $\NH$ for the fits that
include a power law since a higher absorption is needed to reduce the flux
from the power law component at these low energies.
This all suggests that here the power law is primarily acting as a substitute
for a soft thermal component and is probably not a realistic description
of the true spectrum of \Velajr.  Nevertheless, there still could be a
non-thermal power law component at higher X-ray energies but at a much
lower flux level than can be determined by the data.

\begin{table*}
\renewcommand{\arraystretch}{1.3}
\centering
\caption{Spectral modelling results using neutron star atmosphere
models \texttt{nsx} and \texttt{nsmaxg} (hydrogen, unless otherwise noted).
Model parameters that are held fixed are NS mass $M=1.4\,\Msun$ and radius
$R=12\mbox{ km}$, $\Remb/R=1$, and distance $d=1.4\mbox{ kpc}$,
unless otherwise noted.
Absorbed 0.5--8~keV flux $f_{0.5-8}^{\rm abs}$ is in units of
$10^{-12}$~erg~cm$^{-2}$~s$^{-1}$, and
power law normalization is in units of $10^{-4}$~cm$^{-2}$~s$^{-1}$~keV$^{-1}$.
Errors are 1$\sigma$.
A constant MOS/pn normalization and a constant ACIS-S/pn normalization are
included in all fits; the best-fit for these two are $0.967\pm0.006$ and
$1.12\pm0.01$, respectively.
For models with \texttt{gabs} components, Table~\ref{tab:lines} gives the
best-fit values for the \texttt{gabs} parameters.}
\label{tab:atm}
\begin{tabular}{ccccccccc}
\hline
Model & $B$ & $\NH$ & $k\Teff$ & $\Rem$ & $k\Teffb$ & & $f_{0.5-8}^{\rm abs}$ & $\chi^2$/dof \\
& ($10^{11}\mbox{ G}$) & ($10^{21}\mbox{ cm$^{-2}$}$) & (eV) & (m) & (eV) & & & \\
\hline
\texttt{nsx} [carbon] & 0 & 6.81$^{+0.06}_{-0.06}$ & 149$^{+1}_{-1}$ & 5190$^{+150}_{-140}$ & & & 1.254$\pm$0.004 & 545/472 \\
\texttt{nsx} & 0 & 5.03$^{+0.07}_{-0.06}$ & 301$^{+2}_{-2}$ & 871$^{+13}_{-10}$ & & & 1.251$\pm$0.004 & 515/472 \\
\texttt{nsmaxg} & 7 & 4.75$^{+0.09}_{-0.07}$ & 307$^{+2}_{-3}$ & 824$^{+18}_{-12}$ & & & 1.246$\pm$0.004 & 516/472 \\
\texttt{nsx}+\texttt{nsx} & 0 & 5.68$^{+0.47}_{-0.38}$ & 298$^{+2}_{-2}$ & 897$^{+17}_{-14}$ & 57$^{+7}_{-9}$ & & 1.251$\pm$0.004 & 511/471 \\
(\texttt{nsx}+\texttt{nsx})$\times$\texttt{gabs} & 0 & 6.70$^{+0.23}_{-0.29}$ & 303$^{+2}_{-3}$ & 862$^{+23}_{-19}$ & 76$^{+3}_{-4}$ & & 1.252$\pm$0.004 & 471/468 \\
(\texttt{nsx}+\texttt{nsx})$\times$\texttt{gabs}$\times$\texttt{gabs} & 0 & 6.15$^{+0.35}_{-0.50}$ & 303$^{+2}_{-3}$ & 859$^{+22}_{-18}$ & 72$^{+4}_{-6}$ & & 1.252$\pm$0.004 & 466/465 \\
\texttt{nsmaxg}+\texttt{nsmaxg} & 7+0.316 & 5.66$^{+0.26}_{-0.29}$ & 303$^{+2}_{-3}$ & 860$^{+18}_{-12}$ & 60$^{+3}_{-5}$ & & 1.246$\pm$0.004 & 503/471 \\
\texttt{nsx}+\texttt{nsx} [$d=1\mbox{ kpc}$] & 0 & 5.67$^{+0.66}_{-0.35}$ & 298$^{+2}_{-3}$ & 645$^{+17}_{-12}$ & 50$^{+8}_{-7}$ & & 1.251$\pm$0.004 & 510/471 \\
\texttt{nsmaxg}+\texttt{nsmaxg} [$d=1\mbox{ kpc}$] & 7+0.316 & 5.62$^{+0.30}_{-0.29}$ & 302$^{+2}_{-3}$ & 622$^{+14}_{-12}$ & 52$^{+3}_{-4}$ & & 1.246$\pm$0.004 & 504/471 \\
\hline
&&&&& $\Gamma$ & PL norm && \\
\hline
\texttt{nsx}+PL & 0 & 8.42$^{+0.64}_{-0.72}$ & 287$^{+2}_{-2}$ & 967$^{+17}_{-14}$ & 4.33$^{+0.22}_{-0.23}$ & 5.6$^{+1.6}_{-1.6}$ & 1.257$\pm$0.004 & 489/470 \\
\texttt{nsmaxg}+PL & 7 & 8.41$^{+0.62}_{-0.73}$ & 285$^{+2}_{-2}$ & 957$^{+19}_{-17}$ & 4.19$^{+0.21}_{-0.23}$ & 6.1$^{+1.6}_{-1.6}$ & 1.257$\pm$0.004 & 485/470 \\
\hline
\end{tabular}
\end{table*}

\begin{figure*}
\includegraphics[width=\columnwidth]{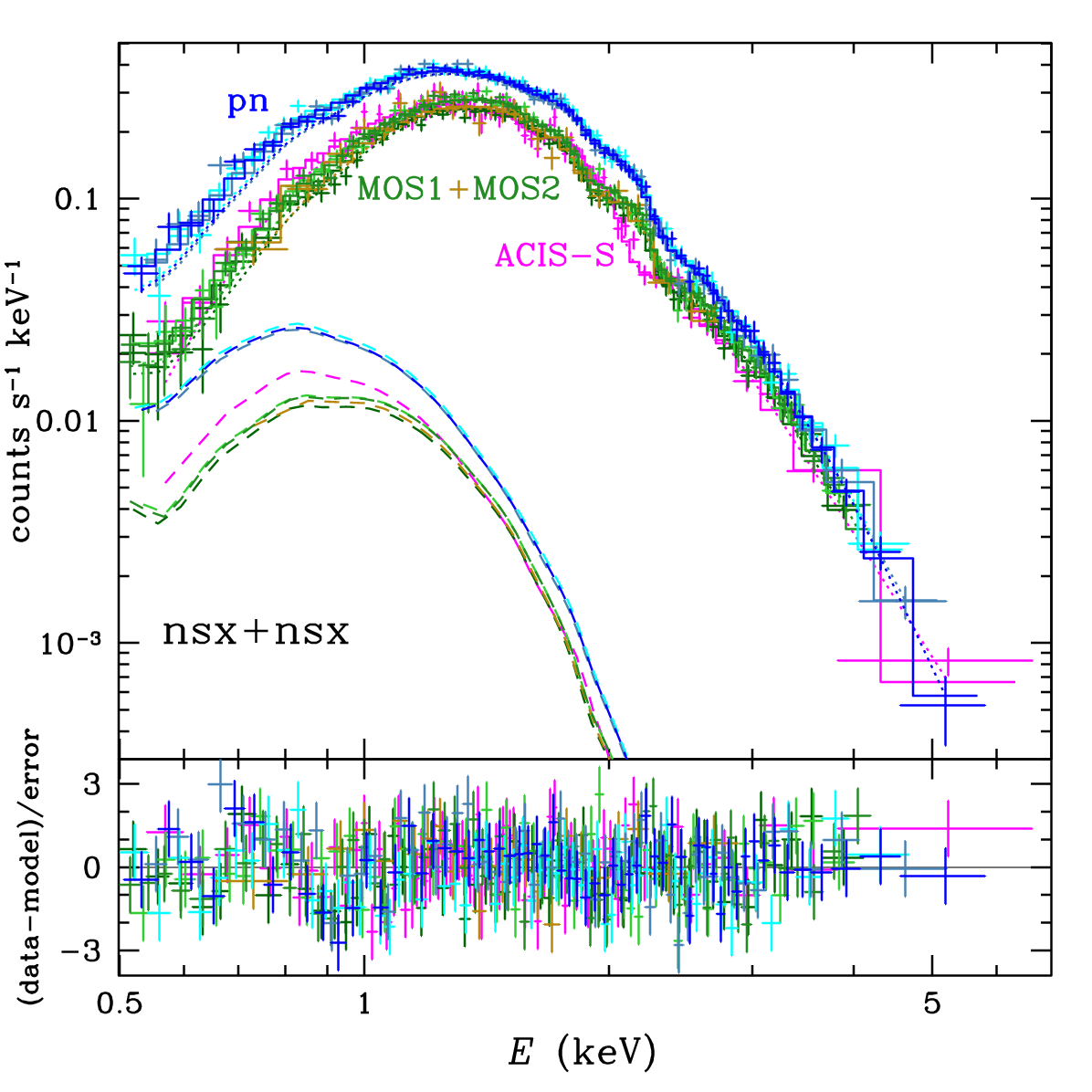}
\hspace{1em}
\includegraphics[width=\columnwidth]{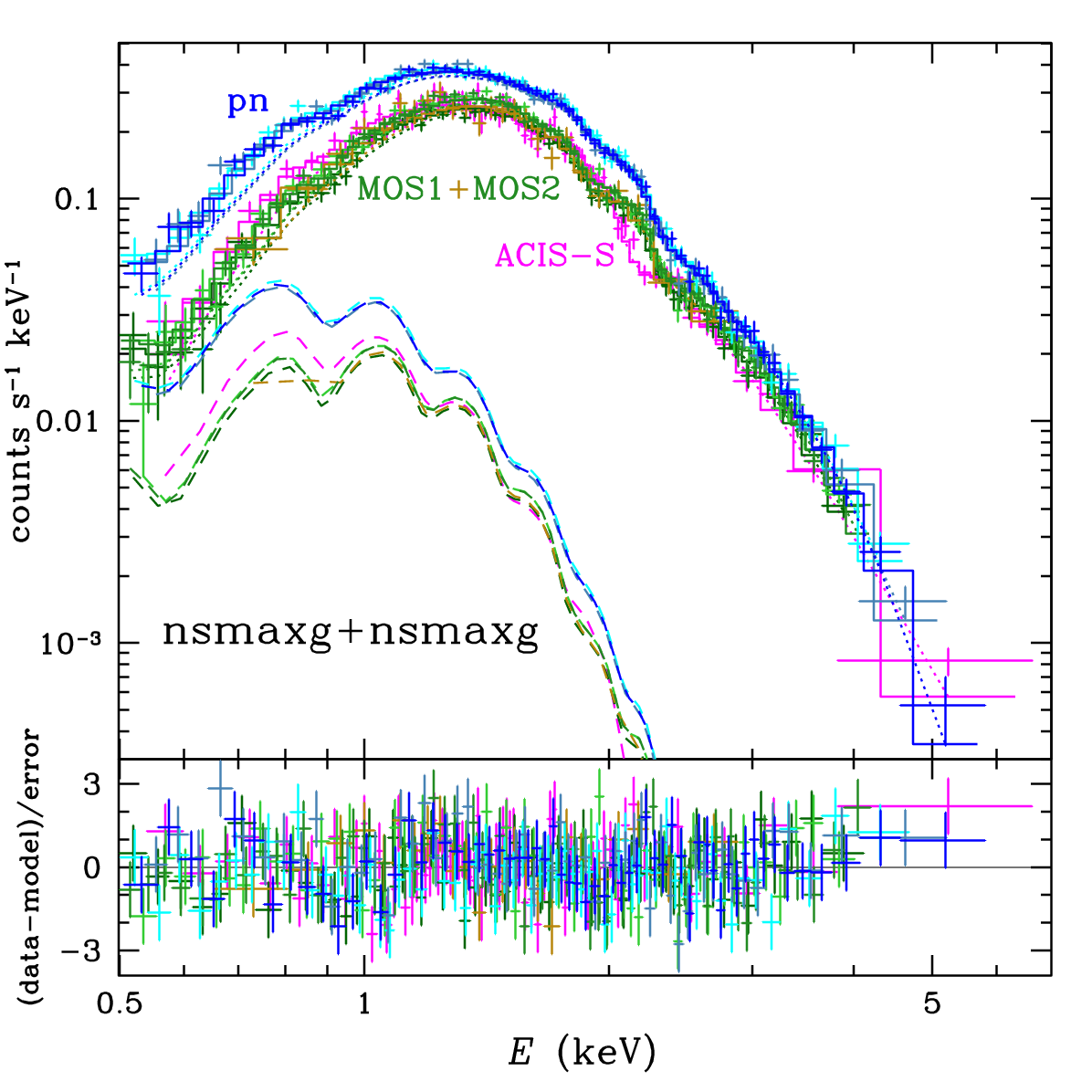}
\caption{
Spectra of \Velajr\ from Chandra ACIS-S and \XMM\ MOS1+MOS2 and pn data.
Top panels show data with 1$\sigma$ errors (crosses) and
spectral model made up of two partially ionized
non-magnetic (\texttt{nsx}+\texttt{nsx}; left)
and magnetic (\texttt{nsmaxg}+\texttt{nsmaxg}; right)
hydrogen atmosphere components
(dashed and dotted lines for cool and hot components, respectively).
For the models on the right, the magnetic fields are
$B=3.16\times10^{10}\mbox{ G}$ and $7\times10^{11}\mbox{ G}$ for the
cool and hot components, respectively.
Bottom panels show $\chi^2=\mbox{(data-model)/error}$.
}
\label{fig:2atm}
\end{figure*}

Next, we fit all the spectra simultaneously using NS atmosphere models,
i.e., non-magnetic \texttt{nsx} and magnetic \texttt{nsmaxg}
(see Section~\ref{sec:models}).
The best-fit results are given in Table~\ref{tab:atm}.
For a single component absorbed atmosphere model, a partially ionized
non-magnetic hydrogen atmosphere provides a satisfactory fit to the data,
with $\chi^2/\mbox{dof}=515/472=1.09$, and is almost as good as a model
using two blackbodies.
Non-magnetic carbon is not as good a fit as hydrogen but better than a
single blackbody.
Keeping in mind the relatively poorer fit we find for a carbon atmosphere,
as well as the fit qualities found previously (see Table~\ref{tab:previous},
in particular, \citealt{potekhinetal20,wuetal21,alfordhalpern23}),
our results are in broad agreement with those from previous works,
except for a somewhat smaller $\Rem$ ($=3.7\mbox{ km}$) when scaled to 1~kpc
and a higher temperature in some comparisons.
For a magnetic hydrogen atmosphere, the two \texttt{nsmaxg} models at
$B=7\times10^{11}\mbox{ G}$ and $9\times10^{11}\mbox{ G}$ provide
about the same quality of fits ($\Delta\chi^2=2$) and parameters that
are within their errors, except for a $\sim 10$~percent difference in
emitting radii.  Poorer fits are obtained at $B\le5\times10^{11}\mbox{ G}$,
with $\Delta\chi^2\ge92$, and at $B\ge10^{12}\mbox{ G}$,
with $\Delta\chi^2\ge10$.

Table~\ref{tab:atm} also gives results for fits using an absorbed
atmosphere and power law model.
We see that such models, with $\Gamma>4$, can fit the data better than
purely thermal models.
However, as we pointed out above, the dominance of the power law component
at low energies, as illustrated in Figure~\ref{fig:nsxpl}, and ensuing
higher $\NH$ argues against these models.

We perform two component absorbed atmosphere fits using the same sets of
models as above.  In this case though, we force the second component to
have a fractional area $\Remb/R=1$, so that this component represents
emission from the whole star at an average temperature of $\Teffb$,
while the first component is due to a small hot spot.
The fit using non-magnetic hydrogen, with $\chi^2/\mbox{dof}=511/471=1.08$,
is a small improvement over the single component atmosphere fits and are
as good as the double blackbody fit.
Our results are broadly in agreement with those of \citet{wuetal21}
(see also Table~\ref{tab:previous}).
Meanwhile, magnetic hydrogen with a small hot spot at
$B=7\times10^{11}\mbox{ G}$ and the remaining cool surface at
$B=3.16\times10^{10}\mbox{ G}$ and $\Teff=6.9\times10^5\mbox{ K}$ gives a
better fit, with $\chi^2/\mbox{dof}=503/471=1.07$, than all the other
purely thermal models tried so far.
Figure~\ref{fig:2atm} shows the partially ionized non-magnetic hydrogen
atmosphere model (\texttt{nsx}+\texttt{nsx}) and magnetic hydrogen
(\texttt{nsmaxg}+\texttt{nsmaxg}) results.
We note for comparison that, if the second (colder) component has a magnetic
field $B=5\times10^{11}\mbox{ G}$, the best-fit is worse by $\Delta\chi^2=8$
and has a somewhat higher $\NH$ and $\Teff$.

While the distance to \Velajr\ was recently pinpointed to
$1.41\pm0.14\mbox{ kpc}$ \citep{suherlietal26},
we also show fit results in Table~\ref{tab:atm} for a distance of 1~kpc.
As expected, the main effect of assuming a smaller distance is a decrease in
the emission radius, so that $\Rem/d$ is approximately constant.
However, there is also an accompanying small decrease in the temperature
$\Teffb$ of the colder surface.

\begin{table}
\renewcommand{\arraystretch}{1.3}
\centering
\caption{Absorption line parameters from using non-magnetic hydrogen
atmosphere models \texttt{nsx}+\texttt{nsx} and one or two Gaussian
absorption line components \texttt{gabs}.
Model parameters that are fixed are the same as given in Table~\ref{tab:atm},
and other best-fit parameters are given there too.
Errors are 1$\sigma$.}
\label{tab:lines}
\begin{tabular}{lcc}
\hline
& One line & Two lines \\
\hline
$E_1$ (eV) & $947^{+18}_{-20}$ & $920^{+22}_{-25}$ \\
Width $\sigma_1$ (eV) & $122^{+27}_{-24}$ & $133^{+28}_{-26}$ \\
Depth $d_1$ (eV) & $41^{+15}_{-11}$ & $55^{+22}_{-19}$ \\
$E_2$ (eV) & & $635^{+14}_{-73}$ \\
$\sigma_2$ (eV) & & $5^{+82}_{-3}$ \\
$d_2$ (eV) & & $<101$ \\
\hline
\end{tabular}
\end{table}

\begin{figure}
\includegraphics[width=\columnwidth]{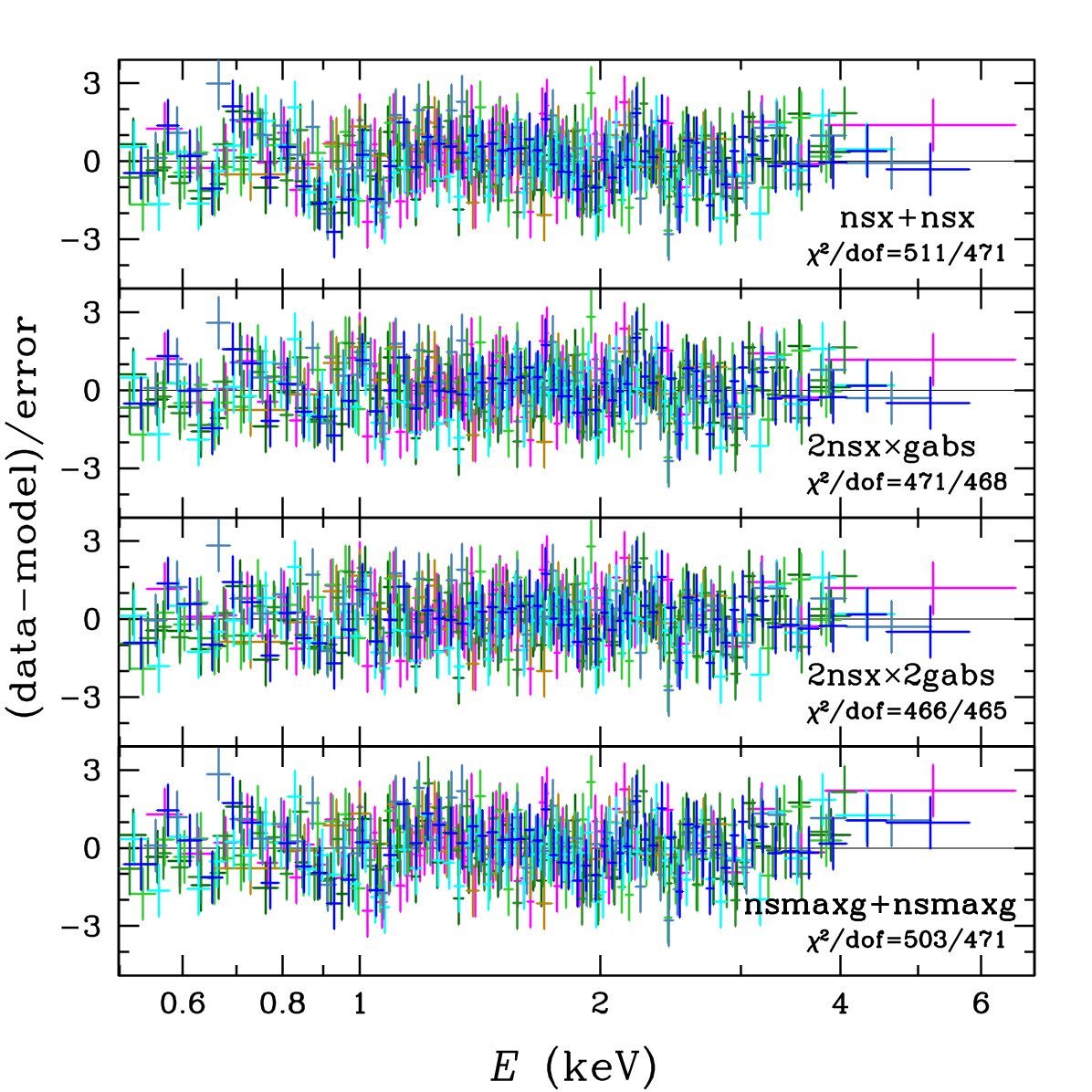}
\caption{
Best-fit residuals using spectral models made up of two partially ionized
hydrogen atmosphere components.
The top panel shows results of non-magnetic \texttt{nsx}+\texttt{nsx},
while the bottom panel shows magnetic \texttt{nsmaxg}+\texttt{nsmaxg} with
$B=3.16\times10^{10}\mbox{ G}$ and $7\times10^{11}\mbox{ G}$ for the
two components (these are the same as in Figure~\ref{fig:2atm}).
The middle two panels show results using \texttt{nsx}+\texttt{nsx} and
including one or two Gaussian absorption lines with \texttt{gabs}.
}
\label{fig:lines}
\end{figure}

The right panel of Figure~\ref{fig:2atm} shows that our best-fit magnetic
hydrogen atmosphere result has spectral features due to transitions at
the harmonics $n$ of the electron cyclotron resonance (redshifted) at
\begin{equation}
\Ebinfty = (\hbar eB/m_{\rm e}c)/(1+\zg)
 = 115.8\mbox{ eV}(B/10^{10}\mbox{ G})/(1+\zg)
\end{equation}
(see, e.g.,
\citealt{pavlovshibanov78,potekhin10,suleimanovetal12,potekhinetal14}).
At $B=3.16\times10^{10}\mbox{ G}$ and $1+\zg=1.235$,
the strongest two lines are at 0.59~keV for the $n=2$ harmonic
and 0.89~keV for the $n=3$ harmonic, while the fundamental
($n=1$) at 0.3~keV is below the energy range we are sensitive to here.
Note also that the best-fit for emission by the dominant hot spot component
has $B=7\times10^{11}\mbox{ G}$, such that $\Ebinfty=6.6\mbox{ keV}$ is
located where the measured flux is low.
It is important to point out that (a) we fixed the NS parameters $M$ and
$R$ such that the redshift factor $1+\zg$ is also fixed and
(b) the redshift factor is
degenerate with magnetic field $B$ in determining the observed cyclotron
energy, such that a 10~percent difference in redshift factor would lead to a
10~percent difference in $B$.
If we allow the radius to be different from 12~km, we find a slightly
better fit ($\Delta\chi^2\approx2$) with a larger radius of 13.5~km
($\pm$1~km), which corresponds to a redshift factor of 1.20.

We also computed spectral models at $B=3.16\times10^{10}\mbox{ G}$ but
with the magnetic field direction $\Theta_B=\pi/2$, i.e., tangential
to the surface rather than normal to it.
We find these spectra lead to a very slightly worse fit ($\Delta\chi^2=1$).
This is in part because, at the best-fit $\Teff$, the harmonics above
the fundamental are somewhat weaker and narrower than when the
atmosphere models have $\Theta_B=0$.
We explore next the strength and broadness of the spectral features.

The predicted resonance features at low energies from the magnetic model
at $B\approx3\times10^{10}\mbox{ G}$ contribute to a better overall fit
compared to the blackbody and non-magnetic models considered above,
in particular by reducing the amplitude of what seems to be systematic
oscillatory behaviour in the fit residuals at $E<1\mbox{ keV}$ seen in
Figure~\ref{fig:bb} and the left panel of Figure~\ref{fig:2atm}.
These oscillations appear to have minima at $\sim$0.6~keV and 0.9~keV.
To better characterize these observed spectral features independent of the
magnetic field, we fit the data using a model composed of
\texttt{nsx}+\texttt{nsx} convolved with either one or two Gaussian
absorption lines modeled by \texttt{gabs}.
The results are given in Tables~\ref{tab:atm} and \ref{tab:lines}, and
residuals of the best-fits are shown in Figure~\ref{fig:lines}.
Most of the best-fit parameters are within the uncertainties of the
\texttt{nsx}+\texttt{nsx} model that does not include the Gaussian
absorption lines, but the fit is greatly improved by the presence of
a line at $\sim$0.92~eV and a second line at 0.6~keV.
A f-test comparison of the fit statistics, $\chi^2/\mbox{dof}=471/468$
versus 511/471, gives a probability of $3\times10^{-8}$, or $\approx5\sigma$,
for one line ($1\times10^{-7}$ for two lines);
note the f-test probability is 0.17 when comparing results of one versus
two lines.
Similar levels of fit improvement are obtained when using instead a minimum
of 25 or 200 counts per bin.

Since cyclotron absorption lines at a single magnetic field are better
described as Lorentzian in shape, rather than Gaussian, we also performed
fits using \texttt{lorabs} instead of \texttt{gabs}.
These fits are slightly worse $\Delta\chi^2<7$ and require larger line
widths.
The better fit by non-magnetic models with broader Gaussian-shaped
absorption lines may be indicating that the magnetic field across most of
the surface varies by a small amount, near that of our magnetic model with
$B=3.16\times10^{10}\mbox{ G}$.

\section{Discussion} \label{sec:discuss}

We presented analysis of all the high-quality archival spectra of
the CCO in the \Velajr\ SNR taken by Chandra and \XMM\ from 2001 to 2010.
Each spectrum is generally well-fit by primarily thermal models, e.g.,
two blackbodies or a blackbody and power law.
The spectral model parameters and flux do not differ significantly
and thus are constant between epochs.
As a result, we fit all the spectra simultaneously to obtain more
robust measurements of the flux and spectral parameters of \Velajr.
We find that more physically-motivated NS atmosphere models can fit
the joint spectra, and these include an atmosphere composed of partially
ionized hydrogen at low ($B<10^9\mbox{ G}$) or moderate
($B\sim10^{12}\mbox{ G}$) magnetic field strengths with a small hot spot
at $\Teff\approx3.5\times10^6\mbox{ K}$.
The remaining NS surface is at a lower temperature
$\Teff\approx(6.6-8.8)\times10^5\mbox{ K}$
(range here spans the range of median $\Teffb$ given for fits at 1.4~kpc
in Table~\ref{tab:atm}, which is larger than their individual uncertainty)
and a lower magnetic field.
There are indications of absorption features at 0.6~keV and 0.9~keV,
which could be due to harmonics of the electron cyclotron resonance
at a magnetic field of $3\times10^{10}\mbox{ G}$.
Such a magnetic field is similar to the fields seen in three other CCOs.
It would be valuable to verify the presence of these absorption features
with a much longer observation or more sensitive instrument.

No matter the spectral model, the measured absorbed 0.5--8~keV flux is
$\approx1.25\times10^{-12}\mbox{ erg cm$^{-2}$ s$^{-1}$}$,
and the unabsorbed total flux is
$\approx2.2\times10^{-12}\mbox{ erg cm$^{-2}$ s$^{-1}$}$.
Therefore, at a distance of $1.4\pm0.1\mbox{ kpc}$ \citep{suherlietal26},
the bolometric luminosity is $(5\pm1)\times10^{32}\mbox{ erg s$^{-1}$}$,
which is mainly due to the hot spot.
In contrast, the approximate average luminosity from the entire 12~km cool
surface is $\Linfty\approx(1.3-4.0)\times10^{32}\mbox{ erg s$^{-1}$}$.
Note that \citet{marinoetal24} derived similar luminosities for their NS
cooling study, despite using a smaller distance to Vela Jr.  This is 
because they only used the ACIS-S CC observation, which yields a higher
flux due to contamination by the SNR and poorer spectral modelling and
fits when only this data is considered (see Section~\ref{sec:results} and
Table~\ref{tab:previous}, as well as discussions in \citealt{beckeretal06}).

The small hot spot that we find for \Velajr\ could lead to X-ray pulsations
at the NS spin period like those seen for three CCOs
\citep{zavlinetal00,gotthelfetal05,gotthelfhalpern09}.
However, searches for pulsations using \XMM\ pn data yielded a pulsed
fraction limit of 5~percent
(\citealt{alfordhalpern23}; see also \citealt{wuetal21}).
Such non-detections thus far could simply be due to an unfavorable
viewing geometry or a hot spot located on the rotation axis.

\begin{figure*}
\includegraphics[width=0.45\linewidth]{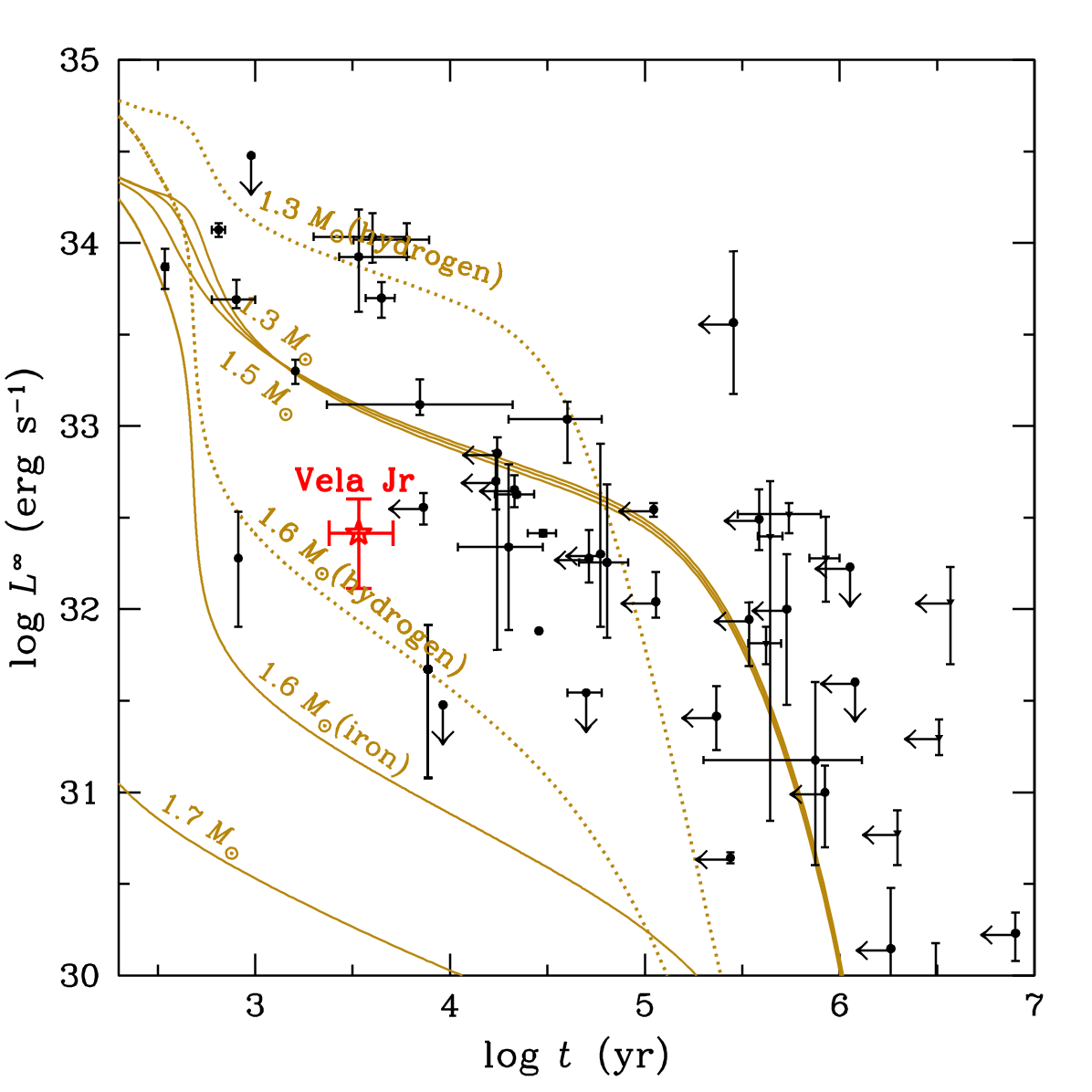}
\hspace{1em}
\includegraphics[width=0.45\linewidth]{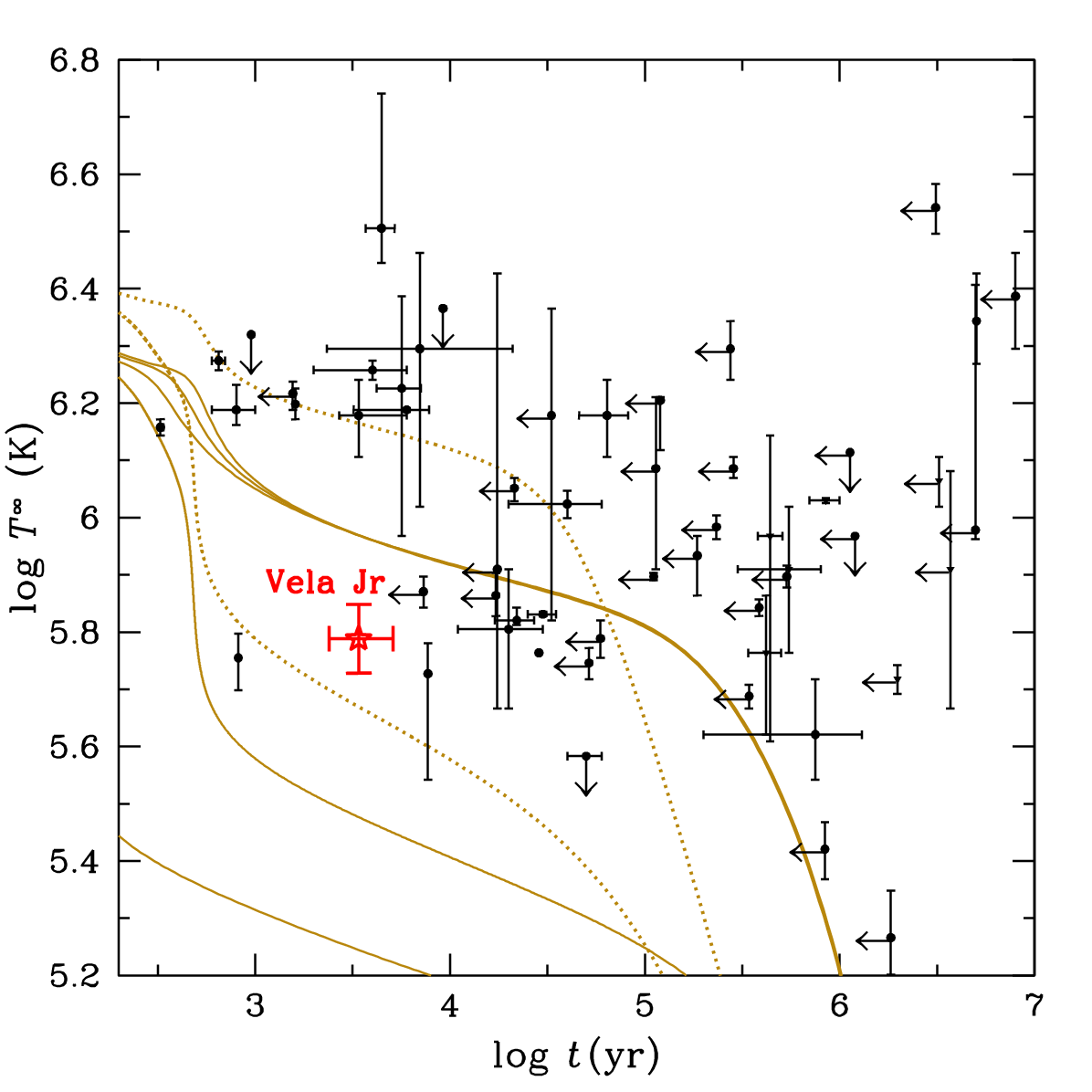}
\caption{Observed luminosity $\Linfty$ (left) and temperature $\Tinfty$
(right) as functions of NS age.
Data points are from \citet{potekhinetal20,hoetal24}
(see also \url{https://www.ioffe.ru/astro/NSG/thermal/}),
except the stars which denote \Velajr.
Solid lines show cooling curves from NS cooling simulations
for NS masses $M=1.3$, 1.4, 1.5, 1.6, and $1.7\,\Msun$
(from top to bottom)
using the BSk24 nuclear equation of state and an iron envelope and
including neutron superfluidity in the crust and core and proton
superconductivity in the core, while dotted lines are for a hydrogen envelope;
note that direct Urca cooling becomes active for BSk24 at $M>1.59\,\Msun$.
See text for more details.}
\label{fig:nscool}
\end{figure*}

Finally, we examine here the impact on NS cooling theory of our
measurements of flux/luminosity and temperature of \Velajr.
In Figure~\ref{fig:nscool}, we plot the luminosity and temperature,
assuming an age of 2.4--5.1~kyr \citep{allenetal15}.
For the luminosity, we use the average value from the entire surface
since the total measured luminosity is dominated by emission from the
small hot spot, as shown by Figure~\ref{fig:2atm}.
Similarly with the temperature, we use the value from the colder
component that represents the temperature of most of the cool surface.
The ranges in $\Linfty$ and $\Tinfty$ use the range of median $\Teffb$ in
Table~\ref{tab:atm}, which is larger than the uncertainty in each $\Teffb$.
Also shown are
the luminosity and temperature of other NSs of various ages
\citep{potekhinetal20,hoetal24}.
The luminosity, temperature, and age of \Velajr\ place it near several
other young NSs, in particular,
PSR~J0205+6449 in the 3C~58 SNR with
$\Linfty\sim2\times10^{32}\mbox{ erg s$^{-1}$}$, $\log\Tinfty\approx5.8$,
and age~$=845\mbox{ yr}$
\citep{slaneetal04,kothes13,potekhinetal20,marinoetal24},
PSR~B2334+61 in the G114.3+0.3 SNR with
$\Linfty\sim5\times10^{31}\mbox{ erg s$^{-1}$}$, $\log\Teff\approx5.6$,
and age~$\sim7.7\mbox{ kyr}$
\citep{yaruyanikeretal04,mcgowanetal06,potekhinetal20,marinoetal24},
and PSR~J0007+7303 in the CTA~1 SNR with
$\Linfty<3\times10^{31}\mbox{erg s$^{-1}$}$, $\log\Teff<6.4$,
and age~$\sim9\mbox{ kyr}$
\citep{caraveoetal10,martinetal16,potekhinetal20}.
\Velajr, these three young NSs, and a few others appear much less
luminous and cooler than most other NSs of similar ages.

Figure~\ref{fig:nscool} also shows the luminosity and temperature evolutions
calculated from NS cooling simulations (see \citealt{hoetal24}, for details).
These simulations use a particular nuclear equation of state, BSk24
\citep{pearsonetal18}, and superconducting and superfluid energy gaps.
But it is important to note that cooling evolutions can vary greatly
depending on which equation of state and energy gaps are used.
Nevertheless, the behaviour, especially regarding direct Urca cooling
and NS mass, is still qualitatively similar despite differences in
theoretical models
(see, e.g., reviews by \citealt{yakovlevpethick04,potekhinetal15}).
For the BSk24 equation of state, it has a threshold mass of $1.59\,\Msun$,
above which the proton fraction in the NS core is high enough that
fast neutrino emission from direct Urca processes is activated.
Meanwhile, some equations of state, such as
BSk26 \citep{pearsonetal18} and SLy4 \citep{douchinhaensel01},
do not have a threshold mass below their predicted maximum mass, and thus
fast direct Urca cooling never occurs for these equations of state.
Below the threshold, NSs cool predominantly by slow neutrino emission
from modified Urca processes, which can be enhanced by Cooper pair breaking and
formation processes \citep{gusakovetal04,pageetal04}.
Thus, we see from Figure~\ref{fig:nscool} that most NSs have a mass
lower than a threshold mass.
On the other hand, colder NSs point to the existence of cooling by
direct Urca processes and hence a nuclear equation of state that allows
for such fast cooling.
The results we derived here more robustly, by using all available
X-ray spectral data and a wide range of spectral models, indicate that,
like PSR~J0205+6449 \citep{pageetal04,slaneetal04,yakovlevpethick04},
PSR~B2334+61 \citep{pageetal09}, and
PSR~J0007+7303 \citep{pageetal04,pageetal09},
\Velajr\ is a strong candidate to be used for this purpose.

\section*{Acknowledgements}

WCGH thanks Peter Shternin for providing the results of his spectral analysis.
The authors thank the anonymous referee for comments that led to improvements
in the paper.
This research made use of data obtained from the Chandra Data
Archive and the Chandra Source Catalog, and software provided by the
Chandra X-ray Center (CXC) in the application packages CIAO and Sherpa.
This work is based on observations obtained with XMM-Newton, an ESA
science mission with instruments and contributions directly funded
by ESA Member States and NASA.
This work uses data and software provided
by the High Energy Astrophysics Science Archive Research Center
(HEASARC), which is a service of the Astrophysics Science Division
at NASA/GSFC and High Energy Astrophysics Division of the
Smithsonian Astrophysical Observatory.
This work made extensive use of the NASA Astrophysics Data System
(ADS) Bibliographic Services and the arXiv.

\section*{Data Availability}

Data underlying this article will be shared on reasonable request
to the corresponding author.
 

\bibliographystyle{mnras}
\bibliography{arxiv}






\bsp
\label{lastpage}
\end{document}